\documentstyle[aps,prl,preprint]{revtex}
\draft
\begin{document}
\author{P.H.M. van Loosdrecht\cite{pvl}, J. Zeman\cite{jz}, G. Martinez}
\address{Grenoble High Magnetic Field Laboratory,
Max-Planck-Institut f\"ur Festk\"orperforschung and
Centre National de la Recherche Scientifique, 25 avenue des Martyrs,
BP 166, F-38042 Grenoble Cedex 9, France}
\author{G. Dhalenne, A. Revcolevschi}
\address{Laboratoire de Chimie des Solides, Univerit\'{e} de Paris-Sud,
b\^atiment 414, F-91405 Orsay, France}
\date{23 Sept. 1996}
\title{Magnetic interactions and the pressure phase-diagram
of CuGeO$_{\rm\bf 3}$.}
\maketitle
\begin{abstract}
A pressure and temperature dependent
Raman study of the vibrational and spin dynamics in CuGeO$_{\rm 3}$\ is
presented.
A new low temperature, high pressure phase has been identified, and
a pressure-temperature phase-diagram is proposed for CuGeO$_{\rm 3}$.
The pressure dependence of the effective exchange interaction, of the
spin-Peierls gap, and of the spin-Peierls temperature strongly supports
a model in which next nearest neighbor interactions stabilise the SP
ground state. The Raman data allow for a quantitative estimate of
the pressure dependence of the next nearest neighbor interactions.
\end{abstract}


The discovery of the first inorganic compound showing the spin-Peierls
(SP) transition, CuGeO$_{\rm 3}$, has sparked an active research into the
physics
and properties of this quasi-one-dimensional magneto-elastic
S=1/2 Heisenberg antiferromagnet \cite{HAS93,BRA83,BOU96,ISO96}.
First evidence for a SP transition in CuGeO$_{\rm 3}$\ came from the
observation of the characteristic exponential vanishing of all
components of the magnetic susceptibility below
$T_{\rm sp}\approx14\ {\rm K}$ \cite{HAS93}. Shortly after, inelastic neutron
scattering (INS) and X-ray diffraction confirmed the SP
transition by the observation of the opening of a gap in the magnetic
excitation spectrum,
and the associated dimerisation of the lattice
\cite{NIS94,HIR94,POU94}. It is now well established that the
($H-T$) phase-diagram of CuGeO$_{\rm 3}$\ conforms to the standard SP
behavior,
{\it i.e.} a uniform phase above T$_{\rm sp}$, a spin-Peierls
or dimerised phase below T$_{\rm sp}$, and a modulated phase at
high magnetic fields ($T<T_{\rm sp}(H))$ \cite{HAS93a}.

Although CuGeO$_{\rm 3}$\ appears to be a good example of a standard SP
system, there
are several problems which remain open at present. One of these
is the observation that above T$_{\rm sp}$\ the magnetic susceptibility
$\chi(T)$\ strongly deviates \cite{HAS93} from
the usual Bonner-Fisher behavior \cite{BON64} expected for a one-dimensional
S=1/2 Heisenberg antiferromagnet. This deviation can have several
origins such as strong spin-phonon interactions, interchain interactions
(about 10 \% of the intrachain interaction \cite{NIS94}), or
next nearest neighbor (nnn) interaction.
For the latter it has
already been shown \cite{CAS95,RIE95} that this may explain
both $\chi(T>T_{\rm sp})$\ as well as the observed dispersion
for the magnetic excitations in the dimerised phase. Moreover, these
studies estimate that the nearest neighbor (nn) interaction is substantially
higher ($J_{nn}\sim150-170 \ {\rm K}$\ \cite{CAS95,RIE95}) than earlier
determinations
from $\chi(T)$\ \cite{HAS93} and INS \cite{NIS94} data ($J_{nn}\sim90-120 \
{\rm K}$),
in good agreement with the result  $J_{nn}\approx180\ {\rm K}$\
obtained from high field Faraday rotation experiments \cite{NOJ95}.
Since a frustrated spin-chain with
$\alpha=J_{nnn}/J_{nn}\geq\alpha_c\sim0.24-0.3$\ \cite{CAS95}
has a spin-gap and a singlet ground state --- even without spin-phonon
interactions --- it is very likely that not
only spin-phonon interactions, but also the nnn interactions in CuGeO$_{\rm
3}$\
play an important role in stabilizing the SP ground state. This is also
in good agreement with recent magnetostriction and thermal
expansion experiments \cite{BUC96}

Another problem which remains open at present is the existence of a
soft mode. Despite serious efforts \cite{LOR94},
the soft mode predicted by standard SP theory \cite{PYT74,CRO79} has so far
not been detected in CuGeO$_{\rm 3}$. Though this might be due to a weak
intensity of
the concerned phonon in INS experiments, its apparent absence
raises serious doubts about the nature of the phase transition itself.

The structure of CuGeO$_{\rm 3}$\ consists of flat CuO$_4$\ and tetrahedral
GeO$_4$\ units, which form chains running along the orthorhombic c-axis
(space group $Pbmm$\ \cite{VOL67}, see Fig. \ref{pressdia}a). The magnetic
chains are formed by the Cu$^{\rm 2+}$\ ions of the CuO$_4$\ units, with a
nn exchange path running over the Cu-O(2)-Cu bonds. The intra-chain
frustration is thought to be due to super-exchange over Cu-O(2)-O(2)-Cu
paths\cite{CAS95}.

In view of the pronounced magneto-elastic properties of CuGeO$_{\rm 3}$, a
study of the vibrational and magnetic excitations under hydrostatic
pressure provides an attractive method to obtain a better understanding of
the interplay between phonons and magnetic excitations in CuGeO$_{\rm 3}$.
Previous experiments have shown that CuGeO$_{\rm 3}$\ undergoes a phase
transition from the orthorhombic $Pbmm$\ phase I at atmospheric pressure to
a, presumably monoclinic, phase II at $\sim$6-7 GPa \cite{ADA91}. More
recently, it has been proposed that this transition might be due to
intercalation of the methanol/ethanol pressure medium \cite{JAY95}, though
subsequent experiments \cite{GON96} have shown this to be unlikely. A
remarkable observation is the huge 20 \% decrease of the b-axis dimension
at the first order I-II phase transition, demonstrating the strong pressure
sensitivity of the CuGeO$_{\rm 3}$\ lattice. In addition, recent
experiments have shown that the magnetic properties of CuGeO$_{\rm 3}$\ are
also quite sensitive to pressure. INS\cite{NIS95a} results have
demonstrated a strong positive pressure coefficient for the SP gap energy,
whereas susceptibility experiments \cite{TAK95} have shown a strong
increase of T$_{\rm sp}$\ upon increasing pressure. This is in good
agreement with earlier predictions \cite{WIN95,WEI95}, but difficult to
understand in terms of the standard SP theories\cite{PYT74,CRO79} and the
structural changes in CuGeO$_{\rm 3}$\ upon increasing
pressure\cite{KAT95}.

The present paper reports on a detailed study of the vibrational and
magnetic properties of CuGeO$_{\rm 3}$\ as a function of pressure and
temperature using Raman spectroscopy. The results obtained on vibrational
scattering are used to propose a pressure-temperature phase-diagram. The
observed behavior of the magnetic scattering gives strong evidence that nnn
interactions in fact drive the SP transition in CuGeO$_{\rm 3}$. It allows
for an evaluation of the pressure dependence of these interactions.

Raman experiments have been performed using a clamp type diamond anvil
cell. The cell has been mounted in a He flow cryostat with a temperature
regulation better than 1K. Figure \ref{spectra} displays some
representative Raman spectra recorded at $T=7$\ K in a (ZZ) geometry at
various pressures between 0 and 7 GPa, using either methanol/ethanol or Ar
as a pressure medium. Here, (ZZ) indicates the polarisation of the incoming
and scattered light parallel to the chains direction. The two modes around
185 cm$^{\rm -1}$\ and 330 cm$^{\rm -1}$, observed in all spectra below 6
GPa, are due to A$_{\rm g}$\ phonons of the orthorhombic structure. The
typical Raman features \cite{KUR94,LOO96,LEM95} of the SP phase (phase
I$_{\rm sp}$) can be observed in the atmospheric pressure spectrum (bottom
curve): the sharp peak at $E_{\rm min}=30\ {\rm cm}^{\rm -1}$, the broad
two magnon maximum at $E_{\rm max}=230\ {\rm cm}^{\rm -1}$, and the
magnetically activated phonon modes at 107 and 370 cm$^{\rm -1}$, where the
former is strongly Fano distorted due to spin-phonon interaction
\cite{LOO96}. Upon increasing pressure the energy of the 30 cm$^{\rm -1}$\
mode, which is observed up to 4.4 GPa, shows a rapid increase, whereas a
slight decrease is found for the two magnon peak energy.

Go\~ni {\it et al.} \cite{GON96} have proposed a ($T,P$) phase-diagram for
CuGeO$_{\rm 3}$, based on Raman experiments in a He filled diamond anvil
cell. Since the results obtained here are slightly different, we start with
a brief discussion of the phases, and phase-diagram observed in this study.
At atmospheric pressure and ambient temperature CuGeO$_{\rm 3}$\ has a
transparent light blue appearance and an orthorhombic crystal structure.
Its first order Raman spectrum shows, as expected, a total of 12 active
modes in the various geometries \cite{DEV94}. The crystals used in this
study, grown by a floating zone technique \cite{REV93}, did not show any
deviation from the selection rules, indicating a good crystal quality. Up
to 6 GPa the crystals remain in the $Pbmm$ phase, and all Raman active
phonon modes show a positive pressure coefficient.

The first order I-II phase transition is evidenced
by the drastic changes in the Raman spectrum
(compare figure \ref{spectra} bottom and top curves).
This new phase, which has a transparent green appearance,
is believed to be monoclinic  \cite{ADA91}, and
is probably accompanied by a cell doubling in at least one direction.
The observation of this phase transition is in good agreement with
earlier results \cite{ADA91,JAY95,GON96}.

In contrast to earlier results \cite{JAY95} we have observed the I-II phase
transition also with Ar as pressure medium.
In this case, however, the phase transition does not occur immediately.
Between 6 and 7 GPa we observed a coexistence of phase II with another
phase which appears dark blue in color.
A comparison of the Raman spectrum
obtained for this phase with earlier results shows that this phase is
in fact phase IIa of ref. \cite{GON96}.
Above 7 GPa the whole sample has transformed to the
light green phase II. This slightly different phase sequence observed
when using Ar as a pressure medium is most likely due to the
huge contraction of the b-axis at the I-II phase transition, which
is evidently less easily achieved in a van der Waals solid as compared
to a liquid pressure medium. The observation of the I-II transition
using Ar, however, does once more demonstrate that phase II
can not be due to intercalation of alcohols \cite{JAY95}.

Go\~ni {\it et al.} reported the existence of a high pressure phase (phase
Ib; $T<180 \ {\rm K}$, $3<P<6$\ GPa\cite{GON96}), characterized by the
appearance of two new lines, at 150 and 300 cm$^{\rm -1}$, in the Raman
spectrum and the disappearance of the SP transition. In samples used in the
present study we could not observe the I-Ib phase transition. Instead a
different phase transition is observed at $T\approx215\ {\rm K}$\ for
$1.6<P<6$\ GPa to a new phase (phase III), characterized by the appearance
of several new lines, at 247, 393, and 471 cm$^{\rm -1}$, in the Raman
spectrum (see arrows in Fig. \ref{spectra}). A remarkable observation is
that the SP transition is apparently not influenced by the I-III
transition, as evidenced by the appearance of all Raman modes of the SP
phase. In fact we did not observe any discontinuities upon going from phase
I$_{\rm sp}$\ to phase III$_{\rm sp}$. This leads to the conclusion that
the I-III phase transition only involves a small distortion of the crystal
structure, in good agreement with the observation of only a few new Raman
lines in phase III. To conclude the discussion on the different phases in
CuGeO$_{\rm 3}$\ figure \ref{pressdia}b proposes a schematic ($P,T$)
phase-diagram for CuGeO$_{\rm 3}$. The major differences with the
phase-diagram proposed in ref. \cite{GON96} are the presence of phase III
(and absence of phase Ib), and the observation of the SP transition up to
at least 4.4 GPa (phase I$_{\rm sp}$\ and III$_{\rm sp}$). We do not
believe that the origin of these differences is due to non-hydrostaticity,
since here, as well as in ref. \cite{GON96}, no evidence of is found this,
at least below 2.5 GPa. But, since the phase sequence is apparently
sensitive to the pressure medium ({\it i.e.} hydrostaticity), it is not
unlikely that the presence of defects may change this sequence.

\begin{figure}
%\begin{center}\includegraphics[width=7cm]{fig3.eps}\end{center}
\caption{
\label{magnp}
Pressure dependence of a) the bound magnon state (symbols) and the twice
the SP gap (dashed line, ref. \protect\cite{NIS95a}) b) the spin-Peierls
transition temperature, c) the energy of the two-magnon maximum
in the Raman spectra,
and d) the frustration parameter $\alpha$\ (see text).
Open symbols refer to experiments using Ar as pressure medium}
\end{figure}
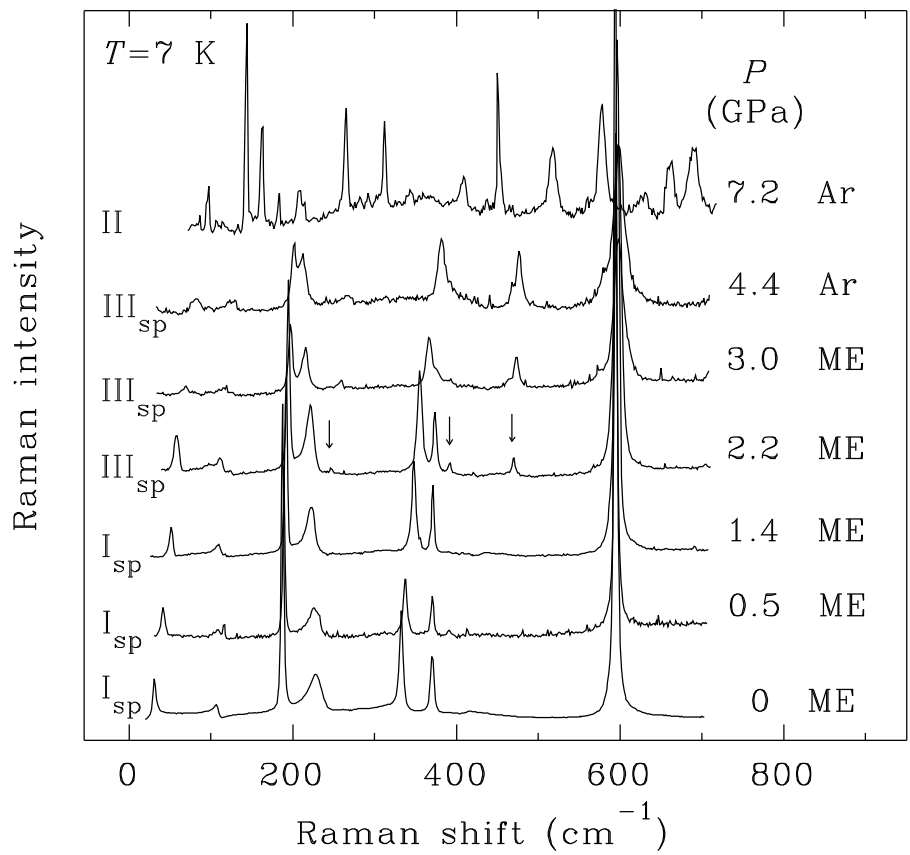
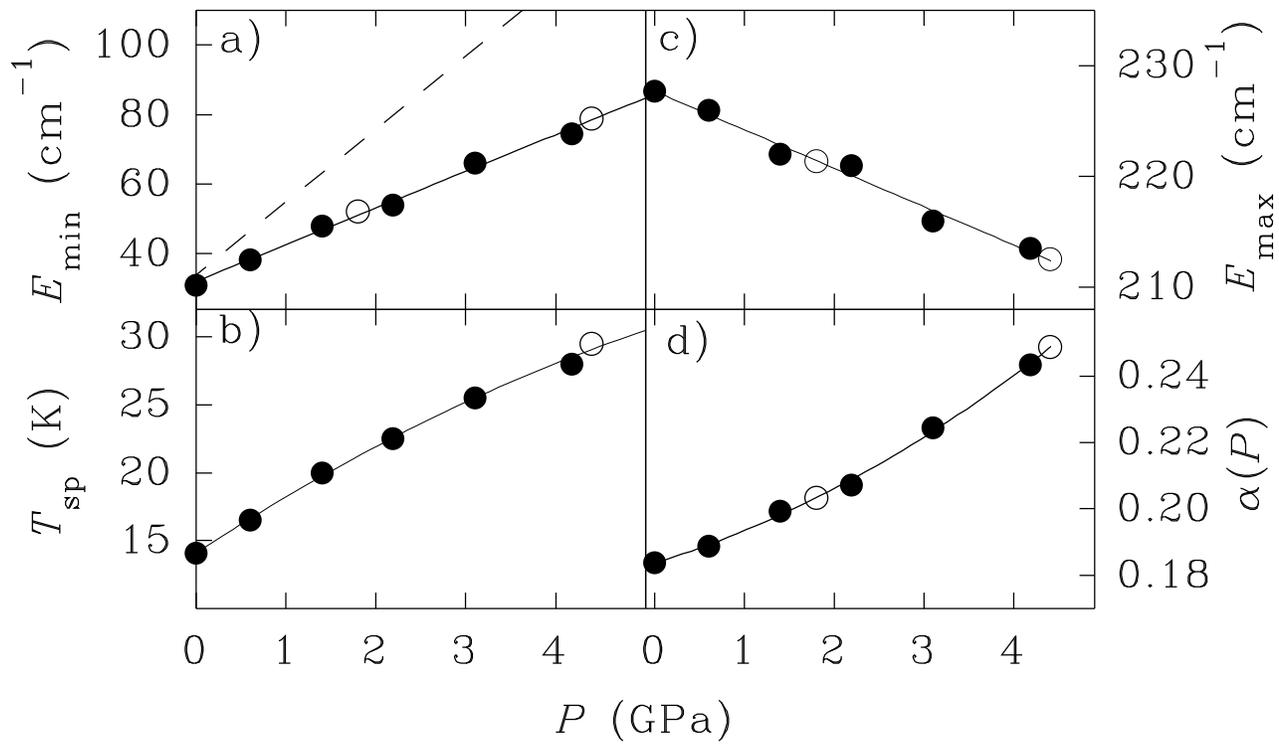
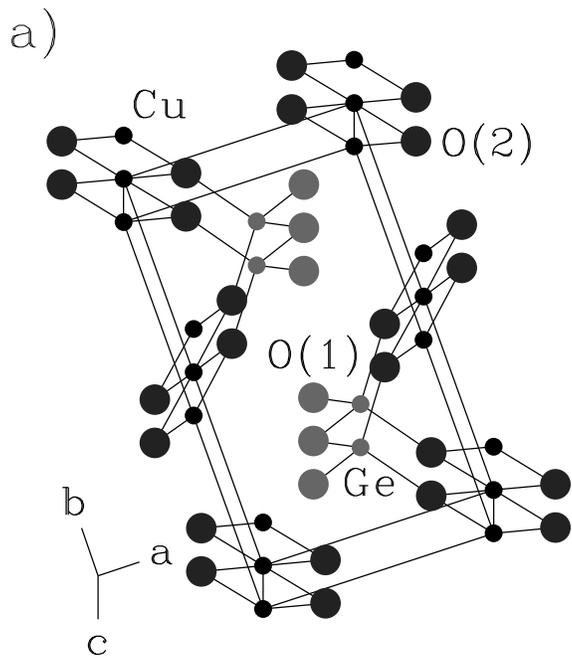
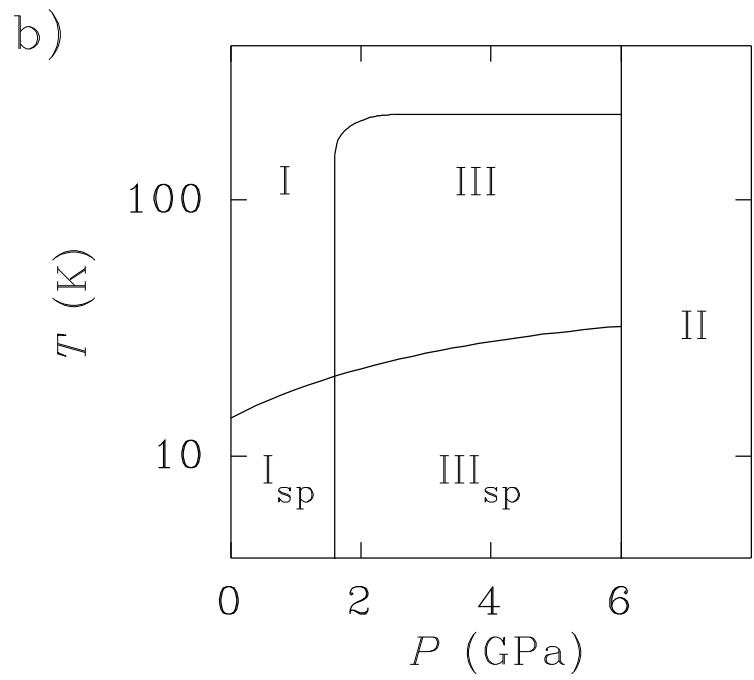

We now turn to the pressure dependence of the magnetic scattering. The
energy of the 30 cm$^{\rm -1}$\ mode is highly sensitive to pressure, as
shown in figure \ref{magnp}a. The straight line through the data
corresponds to $E_{\rm min}[{\rm cm}^{\rm -1}]=31.6+0.11P[GPa]$. For
comparison figure \ref{magnp}a also shows the pressure dependence of the
twice the SP gap as derived from INS experiments \cite{NIS95a}. Clearly the
Raman mode does not correspond to twice the energy gap, as has been assumed
earlier \cite{LOO96,LEM95,LOA96}. Rather, we believe that this deviation
gives clear evidence for the existence of two-magnon bound
state\cite{KUR94}, with a strongly pressure dependent binding energy
($\sim2{\rm cm}^{\rm -1}$\ at 0 GPa, and $\sim15{\rm cm}^{\rm -1}$\ at 4.4
GPa). This conclusion limits the interpretation of the Raman spectrum in
terms of frustration alone\cite{MUT96}. Based on the present results one
may say that the magnetic Raman response in the SP phase is a superposition
of this bound state and a weighted two- or three-dimensional spin density
of states \cite{KUR94,LOO96,LEM95}. The exact nature of the scattering
mechanism, however, remains unclear at present. The usual exchange
scattering mechanism \cite{FLE68} is unable to explain the observed
magnetic Raman activity \cite{MUT96}, indicating that additional
mechanisms, such as for instance spin-phonon interactions, need to be
included to obtain a full understanding of the Raman spectrum of
CuGeO$_{\rm 3}$.

The presence of the bound state mode can be used to determine the
spin-Peierls transition temperature. The pressure dependence of T$_{\rm
sp}$\ obtained in this way is plotted in figure \ref{magnp}b. We find a
much more linear behavior than reported previously \cite{GON96}: $T_{\rm
sp}=14+4.5\ P -0.25 P^{\rm 2}$\ (figure \ref{magnp}b, solid line). The
pressure dependence of T$_{\rm sp}$\ at low pressures is found to be in
good agreement with the results from INS and susceptibility experiments
\cite{NIS95a,TAK95}, as well as with predictions from thermodynamic
experiments \cite{WIN95,WEI95}.

The peak around $E_{\rm max}=230$\ cm$^{\rm -1}$, and the decreasing
continuum towards lower energies, originates from two-magnon exchange
interaction scattering processes and reflects the magnon density of
states\cite{KUR94,LOO96,FLE68}. In systems without frustration, $E_{\rm
max}$ gives an estimate of the effective nn exchange energy, provided one
makes a first order correction to the energy scale in order to account for
magnon-magnon interactions ($E_{\rm max}=2.7 J_{nn}$) \cite{LYO88}. In a
frustrated system, however, one should replace $J_{nn}$\ by an effective
interaction $J_{\rm eff}$, depending both on $J_{nn}$, and $J_{nnn}$. Raman
data at atmospheric pressure yield $J_{\rm eff}=121$\ K (10.4 meV), in good
agreement with INS data \cite{NIS94}. The pressure dependence of $E_{\rm
max}$\ is plotted in figure \ref{magnp}c. Though a clear decrease is
observed upon increasing pressure, it is much weaker than predicted from
INS data\cite{NIS95a}. This deviation may result from a pressure dependence
of the ratio $\alpha=J_{nnn}/J_{nn}$, which has not been considered in ref.
\cite{NIS95a}. The pressure dependent structural changes observed in
CuGeO$_{\rm 3}$\ \cite{KAT95} mainly involve the bonds in the $b$-direction
(Ge-O(2), Cu-O(1)), which presumably play an important role in the nnn
interactions. According to ref. \cite{KAT95}, the changes in the length and
angle of the O(2)-Cu-O(2) bonds are quite weak. One therefore expects only
a weak pressure dependence of $J_{nn}$, and we are led to assume that the
main pressure variation of $\alpha$\ is induced by that of $J_{nnn}$.

Using a simple model, one can derive the pressure dependence of $\alpha$\
from the position of the two-magnon maximum.
The hamiltonian describing a dimerised linear frustrated spin chain can be
expressed as
\begin{equation}
H=\sum_i J_{nn} (1+\delta (-1)^i)S_iS_{i+1}+J_{nnn} S_iS_{i+2}\ \ ,
\end{equation}
where $\delta$\ is due to the dimerisation.
Without dimerisation, linear spin-wave theory gives a dispersion
relation $E_{\delta=0}(q)=\zeta(q,\alpha)E_0\sin(q)$, with
$\zeta(q,\alpha)=\sqrt{(1-4\alpha(1-\alpha\sin(q))}$, $E_0=(\pi/2)J_{nn}$\
the maximum spin excitation energy for $\alpha=0$, and $q$\ the wavevector
along the chains. The SP gap $\Delta$\ may be introduced phenomenologically
by writing
\begin{equation}
\label{disp} E^2(q)=\Delta^2 + (\zeta E_0\sin(q))^2
\end{equation}
In order to estimate $\Delta$, $E_0$\  and $\alpha$\ at zero pressure
we have fitted eq. \ref{disp} to INS data \cite{NIS94,REG96}.
A good fit is obtained for $\Delta=2.1$\ meV, $E_0=24.5$\ meV
($J_{nn}=181\ {\rm K}$), and $\alpha=0.18$.
Using eq. \ref{disp}, and the pressure dependence $\Delta(P)$\
obtained in \cite{NIS95a} one may now evaluate $\alpha(P)$.
The resulting pressure dependence $\alpha(P)$, plotted in figure
\ref{magnp}d, clearly show a pronounced increase of the nnn interactions upon
increasing pressure.
In obtaining this result the possible pressure dependence of the nn
interactions ({\it i.e.} $E_{\rm 0}$) has been neglected.
However from structural considerations one would in fact expect a small
{\it increase} of $J_{nn}$\ upon increasing pressure,
which would lead to an even stronger pressure dependence of $J_{nnn}$.
We believe that the observed pressure dependence of $J_{nnn}$\ is due to
the changes in the strong Ge-O(2) bonds \cite{KAT95}, which determine the
orientation of the oxygen ligands, and hence of the Cu-O(2)-O(2)-Cu
super-exchange.

In conclusion, based on pressure dependent Raman data we proposed a ($P-T$)
phase-diagram for CuGeO$_{\rm 3}$\ which differs from earlier
results\cite{GON96}. In particular a new phase III has been observed which
like phase I, also undergoes a SP transition at low temperatures. The
comparison of the pressure dependence of the 30 cm$^{\rm -1}$\ Raman mode
with that of the SP gap clearly shows that the 30 cm$^{\rm -1}$\ mode does
not reflect the density of states. This leads us to conclude that this mode
is due to a two magnon bound state, which, given the strong pressure
dependence of its binding energy, as well as its absence in the uniform
phase, is likely induced by the magnetic dimerisation. Finally, we have
clearly shown that the increase of the SP gap and the SP transition
temperature, as well as the decrease of the maximum energy of the spin
excitations ({\it i.e.} $J_{\rm eff}$) upon increasing pressure has a
natural explanation in terms of nnn interactions. From the present results
one may conclude that the phonons, though essential for the lattice
dimerisation, play only a minor role in determining the SP gap and
transition temperature in CuGeO$_{\rm 3}$, but that these are largely
determined by the ratio of the nn and nnn interactions.

The Grenoble High Magnetic Field Laboratory is a ``laboratoire
conventionn\'e \`a l'Universit\'e Joseph Fourier de Grenoble''

\begin{figure}
\begin{center}
\end{center}
\caption{
\label{pressdia}
a) Crystal structure of CuGeO$_{\rm 3}$.
b) Proposed schematic ($P-T)$ phase-diagram of CuGeO$_{\rm 3}$.}
\end{figure}
\begin{figure}
%\begin{center}\includegraphics[width=7cm]{fig2.eps}\end{center}
\caption{
\label{spectra}
Partially polarised (ZZ) Raman spectra of CuGeO$_{\rm 3}$\ at $T=7$\ K for
various
pressures. The pressure media (ME: methanol/ethanol; Ar: Argon)
used are indicated on the right side, and the different phases on the left
side.}
\end{figure}
\end{document}